\documentclass[apj]{emulateapj}

\usepackage{graphicx}
\def\mr{\mathrm}

\def\kms{\mr{~km\ s}^{-1} }

\def\kmsmpc{\mr{\ km}\ \mr{s}^{-1}\ \mr{Mpc}^{-1}}

\def\mpc{\mr{~Mpc}}

\def\cmc{\ \mr{cm^{-3}}}
\def\cms{\ \mr{cm^{-2}}}
\def\gcmc{\ \mr{g \cmc}}
\def\nfirst{$4\pm1_{-2}^{+6}\times10^{-3}$}
\def\nfirstg{$4\pm1_{-2}^{+6} \times10^{-27} \gcmc\ $}
\def\nsdss{$9\pm3_{-5}^{+10}  \times10^{-4}$}
\def\nsdssg{$9\pm3_{-5}^{+10}  \times10^{-28} \gcmc\ $}
\begin{document}
\title{Bent-Double Radio Sources as Probes of Intergalactic Gas}
\author{E. Freeland, R. F. Cardoso, E. Wilcots}
\affil{Department of Astronomy, University of Wisconsin, Madison, WI 53706}
\email{freeland@astro.wisc.edu, cardoso@astro.wisc.edu, ewilcots@astro.wisc.edu}

\begin{abstract} As the most common environment in the universe, groups of
galaxies are likely to contain a significant fraction of the missing baryons in
the form of intergalactic gas.  The density of this gas is an important factor
in whether ram pressure stripping and strangulation affect the evolution of
galaxies in these systems.  We present a method for measuring the density of
intergalactic gas using bent-double radio sources that is independent of
temperature, making it complementary to current absorption line measurements.
We use this method to probe intergalactic gas in two different environments:
inside a small group of galaxies as well as outside of a larger group at a $2
\mpc$ radius and measure total gas densities of \nfirst $\cmc$ and \nsdss
$\cmc$ (random and systematic errors) respectively.  We use X-ray data to place
an upper limit of $2 \times 10^6$ K on the temperature of the intragroup gas in
the small group.  \end {abstract}


\section{Introduction} In the local universe, attempts to measure the baryon
content in the form of stars, hot X-ray emitting gas, and cold gas account for
only one third of the baryon density seen at high redshift
\citep{2004ApJ...616..643F}.  Simulations predict that the remaining ``missing
baryons" exist in a warm-hot intergalactic medium (WHIM) that was shock heated
during large-scale structure formation to a temperature range of $10^5 < T <
10^7$ K \citep{2006ApJ...650..560C,2001ApJ...552..473D}.  Observational
confirmation of the WHIM comes primarily from UV absorption lines in the
spectra of low redshift quasars \citep{2000ApJ...534L...1T}.  However, UV
absorption line observations are only able to probe a limited temperature range
of WHIM gas and there is continued debate as to whether the absorption
originates in the extended gaseous halos of individual galaxies, the
intergalactic medium (IGM), or from large-scale filaments
\citep{2003ApJS..146..125S}.  

Galaxy groups, the most common environment in the local universe, likely
contain a significant fraction of the total baryonic mass within their
intragroup medium \citep{1998ApJ...503..518F}.  Until now the only probes of
this gas have been UV absorption line and X-ray observations.  In the few
groups that have a hot intragroup medium X-ray observations can determine the
emission measure, but without some knowledge of the geometry the density of
this gas cannot be disentangled.  The X-ray surface brightness typically 
falls below detectable levels at radii much smaller than the
virial radius, while the implied gas mass is still increasing at least linearly
with radius \citep{2000ARA&A..38..289M}.  Groups with a bright quasar behind
them have been observed in an attempt to detect absorption from the intragroup
medium.  In cases where intervening gas is detected, assumptions about the
extent of the absorbing system, the metallicity, ionization fraction, and the
intensity of the ionizing radiation field lead to estimates of the
intragroup medium density on the order of
$n\sim10^{-4}-10^{-5}~\mathrm{cm}^{-3}$
\citep{2004AJ....127..199P,1998ASPC..143..261T}. 

Additionally, galaxy groups, according to the hierarchical scenario of the
formation of large-scale structure, are the building blocks of rich clusters of
galaxies.  Groups are important sites in which to investigate the physical
mechanisms responsible for the observed morphology and SFR-density relations
\citep{2003MNRAS.346..601G,2003ApJ...584..210G}.  However, the effectiveness of
mechanisms like ram pressure stripping and strangulation on galaxy evolution in
the group environment are not well understood due to the uncertainty in the
density of the intragroup medium.  Some observational evidence does exist for
the stripping of hot gas from galaxies by the intragroup medium,
mainly in features like X-ray shocks and tails
\citep{2004ApJ...617..262S,2006MNRAS.370..453R},  as well as HI observations
that show compression features opposite extensions in cold disk gas
\citep{2005A&A...435..483K,2007ApJ...671L..33M}. 
 
Another way to measure the density of the IGM is through its effect on the jets
of radio galaxies.  As a double lobed radio source travels through
intergalactic gas ram pressure causes its jets to be swept back.  Prior to
\citet{1987AJ.....94..587B} it was believed that the combination of the dense
intracluster medium and large galaxy velocities needed to form these
bent-double radio sources could only be found in rich clusters.  While many
bent-doubles do reside in dense clusters, a surprising number are found in
lower mass environments like groups of galaxies
\citep{1978A&A....69..253E,1994ApJ...436...67V,1995AJ....110...46D,2001AJ....121.2915B}.

We present measurements of the density of intergalactic gas from
radio, optical, and X-ray observations of two bent-double radio sources and
their surrounding environments.  Both sources were identified by
\citet{2001AJ....121.2915B} as existing in environments with richness less
than Abell class 0.  
We adopt $H_0=73 \kmsmpc$.

\section{Data} 
\subsection{Radio Observations} We observed the FRI radio
galaxies FIRST J$124942.2+303838$ and SDSS J$085331.86+232400.0$ (hereafter S1
and S2 respectively) in March of 2006 with the Giant Metrewave Radio Telescope
(GMRT).  The GMRT consists of thirty 45 meter diameter antennas with half
arranged in a compact 1 km diameter, randomly distributed, central square and
the remaining in three extended arms with fixed baselines ranging from 0.1 to
25 km in length.  The observations were dual frequency 610/235 MHz using both
the upper and lower sidebands.  In this paper we will discuss only the single
polarization 610 MHz data which have an effective bandwidth of 12 MHz in each
sideband.  We observed a flux calibrator (3C286 or 3C48) at the beginning and
end of each observation and every 40 minutes we observed a phase calibrator for
6 minutes.  We have a total of 5.5 hours integration on each source.The data
were calibrated in AIPS using standard tasks.  Imaging and self-calibration
were done with multiple facets across the $0.7^{\circ}$ (half-power beam width)
field of view.      

\subsection{Optical Observations} Multi-object spectroscopy was performed in
May 2007 and April 2008 using HYDRA on the WIYN 3.5m telescope on Kitt Peak for
the source S1.  Fibers from the blue cables were placed
on galaxies chosen from Sloan Digital Sky Survey (SDSS) photometric redshifts
to have a redshift similar to the radio source, and additional fibers were
placed on blank sky positions.  The 600@10 Zepf grating was used in first order
to give spectra with a dispersion of $1.4$ \AA\ per pixel covering a wavelength
range from 4600 to 7200 \AA.  CuAr lamp calibration spectra were used for the
wavelength calibration and an averaged sky spectrum was subtracted from each
object spectrum.   Spectra were obtained for $\sim 60$ galaxies overall.  The
data were reduced in IRAF using the {\it dohydra} package.  Cross correlation
of object spectra with galaxy template spectra was performed using the RVSAO
\citep{1998PASP..110..934K} package in IRAF.

\subsection{X-ray Observations} We obtained archival Chandra observations of
both sources taken originally in January and February 2005 (PI Blanton).
Exposure times were 35.17 and 47.19 ksec for S1 and
S2, respectively, and both sources were centered on the
back illuminated ACIS-S3 chip.  

The event files were filtered to include events with energies between 0.3 keV
and 7.0 keV.  Point sources were identified using task \emph{wavdetect} and
removed.  Images were made by smoothing the data in two dimensions with a
gaussian extending to 5$\sigma$ and with a full-width at half maximum of 6
pixels on both axes.  Images were also made using an adaptive smoothing with a
minimal significance signal to noise ratio of 3.  Spectra were extracted from a
region $300$ kpc in radius around the radio source as well as a background
region.  These sources are distant enough that they fit on the back illuminated
ACIS-S3 chip with area left for a generous background region.

\section{Method} 
\begin{figure}
\includegraphics[width=\textwidth]{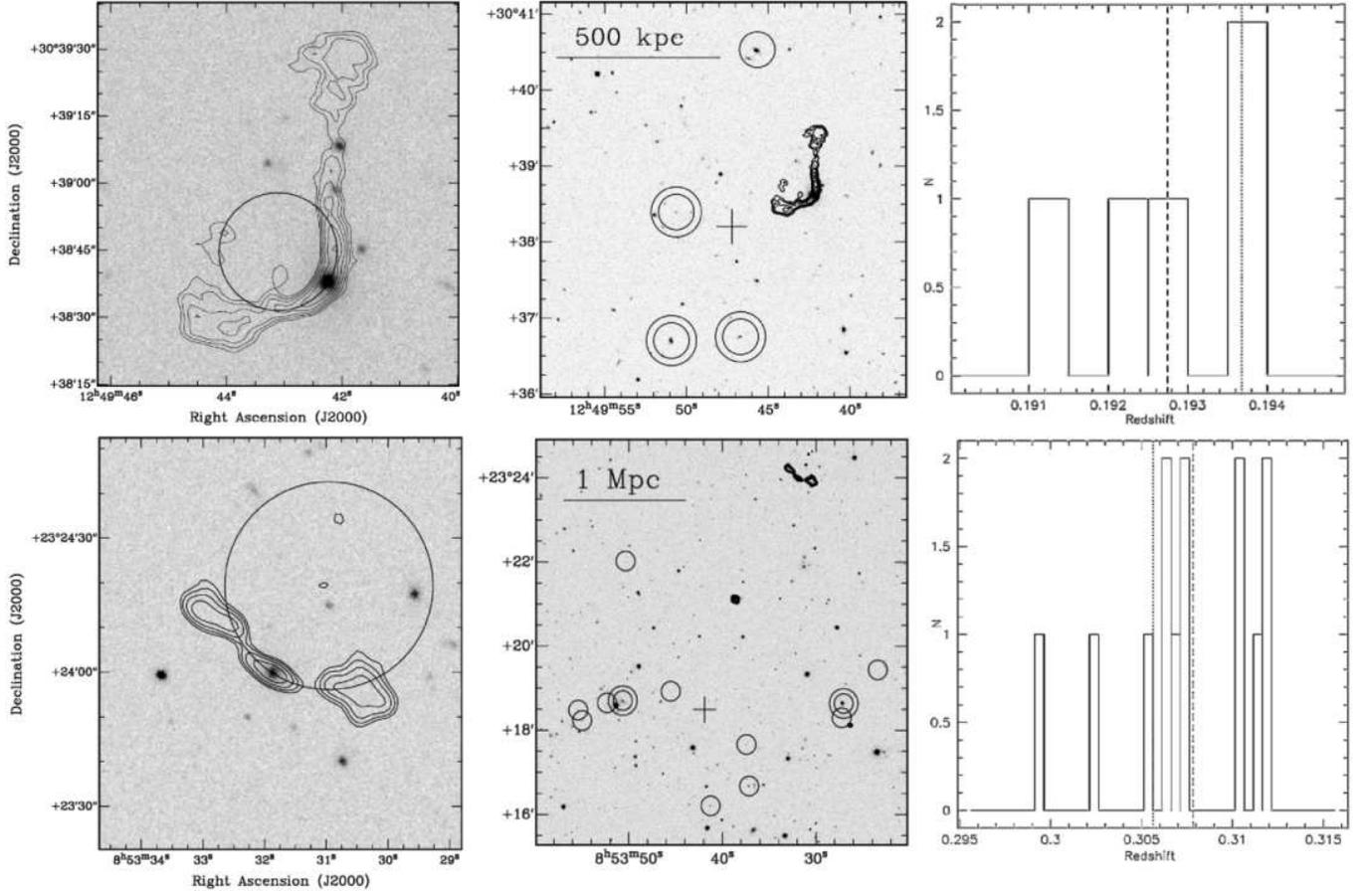}
\caption{610 MHz GMRT radio contours overlaid on SDSS r band optical images, S1
(top) and S2 (bottom).  Lowest contour levels are 1 $\mathrm{mJy\ beam^{-1}}$
(S1) and 1.3 $\mathrm{mJy\ beam^{-1}}$ (S2) and increase by $\sqrt{2}$.  Beam
sizes are $4.3\arcsec \times 3.3\arcsec$ for S1 and $5.8\arcsec \times
4.9\arcsec$ for S2.  In the panels on the left the radius of curvature is
illustrated.  In the center panels, crosses indicate the averaged group center
and galaxies with spectroscopic redshifts are circled twice while photometric
redshifts are circled once.  To the right, the redshift histograms have a
dashed line indicating the group redshift and a dotted line at the redshift of
the bent-double radio source.}  \label{fig:skydist}

\end{figure}

For a relativistic, hydrodynamic jet, the
time-independent Euler equation describes the balance of internal and external
pressure gradients \citep{1979Natur.279..770B,1979ApJ...234..818J,1980AJ.....85..204B},  

\begin{equation} \frac{\rho_{\mbox{\tiny IGM}}
v_{gal}^2}{h}=\frac{\mathrm{w}\Gamma^2\beta^2}{R} \end{equation}  where
$\rho_{\mbox{\tiny IGM}} v_{gal}^2$ is the external ram pressure felt by the
radio galaxy as it travels through the IGM, $\mathrm{w}\Gamma^2\beta^2$ is the
relativistic enthalpy density inside the jet, $h$ is the width of the jet, and
$R$ is the radius of curvature of the jet.  In our data the jet widths are not
resolved and are set by the beam size of the individual observations.  The
radius of curvature, $R$, is found by fitting a circle by eye along the jets
through the core.  This radius is illustrated in the left panels of Figure
\ref{fig:skydist}.  The velocity of the radio galaxy, $v_{gal}$, is estimated
using the velocity dispersion of the group as $\sqrt{3}\sigma_{v}$.  The
enthalpy density is written as $\mathrm{w}=e+p$, where $e$ is the energy
density and $p$ is the internal pressure.  For the jets whose particle
population is ultrarelativistic, $e=3p+\rho_{jet}c^2$ and thus
$\mathrm{w}=4P_{min} + \rho_{jet}c^2$ assuming that the internal pressure in
the jets is dominated by the minimum synchrotron pressure, $P_{min}$.  While
$\rho_{jet}c^2$ is not well known, it is anticipated to be much less than
$4P_{min}$ and we do not include it in our calculations
\citep{1984ARA&A..22..319B}.  We adopt $\beta=v/c=0.54\pm0.18$ as the
distribution of observed jet speeds in FRI radio galaxies with straight jets
\citep{2004MNRAS.351..727A} which is consistent with speeds in wide-angle
tailed radio galaxies \citep{2006MNRAS.368..609J}.  The
relativistic factor $\Gamma$ is the usual $({1-\beta^2})^{-1/2}$.   

The internal pressure is calculated using the minimum synchrotron pressure as
outlined in \citet{1987ApJ...316...95O}.
\begin{equation}P_{\mr{min}}=(2\pi)^{-\frac{3}{7}}\left( \frac{7}{12} \right)
[c_{12}L_{\mr{rad}}(1+k)(\phi V)^{-1}]^{\frac{4}{7}} \ \ \mr{ergs \ cm^{-3}}
\end{equation} where $c_{12}$ is a constant that depends on the spectral index
and frequency cutoffs \citep{1970ranp.book.....P}, $k$ is the ratio of
relativistic proton to relativistic electron energy, $\phi$ is the volume
filling factor, $V$ the source volume, and $L_{\mr{rad}}$ the radio luminosity
of the jet.  We measure this internal pressure at the position immediately
before the jet bends, always excluding the core.  Standard equipartition of
energy between relativistic particles and magnetic fields is assumed.  We
follow the general assumptions that $k=1$, $\phi=1$, and the jets are cylinders
uniformly filled with magnetic fields and relativistic particles.  We use VLA
FIRST survey \citep{1995ApJ...450..559B} data at 1420 MHz and our GMRT 610 MHz
data to determine the spectral index of the synchrotron emission.  Our values
for the minimum synchrotron pressure in the jets are in good agreement with
other measurements for similar radio sources
\citep{1994ApJ...436...67V,2000ApJ...530..719W}. 
 
\section{Group Redshifts and Velocity Dispersions}

In order to estimate the velocity of the radio galaxy through the IGM we use
the optical data to examine the environments of these sources.  Small group
memberships and instrumentation limit our ability to obtain large radial
velocity samples.  Additionally, structures associated with these radio sources
may not have achieved the dynamical equilibrium that would result in a Gaussian
distribution of radial velocities among members.  With these considerations in
mind we use the robust statistical biweight estimators for location and scale
\citep{1990AJ....100...32B} to determine the group redshift and velocity
dispersion for each sample.  These estimators do not assume a Gaussian parent
population and are not strongly influenced by outliers which is especially
important for small samples. 

We combined redshift information from our WIYN spectroscopy (only for S1), as
well as spectroscopic and photometric redshifts from the Sloan Digital Sky
Survey DR6 \citep{2008ApJS..175..297A}.  We use only photometric redshifts
whose errors are less than 5\% of their value for S1 and 10\% for S2.

We calculate a peculiar velocity for a galaxy with redshift $z$ in the rest
frame of the group using \begin{equation} v_{pec} =
c(z-z_{group})/(1+z_{group}).  \end{equation}  The velocity dispersion,
$\sigma_{group}$, will be taken as the dispersion of the $v_{pec}$ values.  We
start by examining the redshift histogram for galaxies within a projected 6 Mpc
distance from the radio source and the sky locations of galaxies within $2000
\kms$.  With so few redshifts available the final samples are chosen by eye
based on the observed spatial clustering seen in Figure \ref{fig:skydist} which
shows the spatial and redshift distributions of galaxies near the two radio
sources.  In the case of S2, the radio source appears offset spatially from the
group.  10 of the galaxies circled in black have photometric redshifts and 3
(including the radio source) are SDSS spectroscopic redshifts.  We estimate the
speed of this source through the IGM using its velocity relative to the mean
redshift of the system it is approaching ($\bar{z}=0.3081$) and find a speed of
$570 \kms$.  The velocity of this source is uncertain although not unreasonable
and we use a rough error of $10\%$ of its value in future calculations.  S1 is
associated with four other nearby galaxies, three have redshifts measured from
our WIYN data and the fourth is a photometric redshift from SDSS.  These five
galaxies have a velocity dispersion of $250_{-110}^{+20} \kms$.  The $68\%$
error bars on the velocity dispersion are determined by generating bootstrap
samples using the final set of $v_{pec}$ values. 
 
\section{IGM Temperature}

We use the X-ray data to examine the temperature of the IGM in these systems.
XSPEC \citep{2004HEAD....8.1629A} is used to fit the spectra extracted from a
region $300$ kpc in radius surrounding the radio source with point sources
removed.  We use Raymond models for thermal X-ray emission from a hot,
optically thin gas, fixing the abundance at Z=0.1 and the normalization
according to the measured number density shown in Table \ref{tab:sour},
assuming it is constant throughout the volume.  All models included a WABS
component of Galactic absorption with values of $1.25 \times10^{20} \cms$ and
$3.23\times10^{20} \cms$ for S1 and S2 respectively, based on the measurements
of \citet{1990ARA&A..28..215D}.   

In the case of S1 the number of counts in the region around the radio source
after background subtraction is only $80\pm35$.  The spectrum from this region
was binned to give each measurement a $3\sigma$ significance level above the
background.  A Raymond model was then compared with the data to determine the
temperature the gas would need to produce counts at a $3\sigma$ level above the
current observations.  Gas with a density of $4\times 10^{-3} \cmc$ and a
temperature higher than $0.2$ keV ($2\times 10^6$ K) should be observable with
the current data.  Thus, we put an upper limit on the temperature of the IGM in
this group at $2\times 10^6$ K.   

After background subtraction we detect $94\pm26$ counts from the source region
for S2.  We do not expect the intergalactic gas around this source to be hot
since it is 2 Mpc from the center of a system of galaxies.  The adaptively
smoothed image shows the X-ray counts concentrated near the radio source.  We
interpret these X-rays as originating from Inverse-Compton scattering off the
relativistic electrons in the radio source.  The data are not adequate to
distinguish between thermal and non-thermal models for the emission.
Additionally, with the current data the density of this gas is too low to
place an upper limit on the temperature were we to assume that these X-rays
originated from a hot IGM. 

\section{Discussion} We present two measurements of the density of
intergalactic gas using radio sources whose jets are bent back by ram
pressure.  S1 is probing gas with a density of \nfirstg (\nfirst $\cmc$
assuming a mean molecular weight of $0.6$ and where the first set of errors
are random and the second systematic) that is near the center of a small group
of galaxies.  In the case of S2 this source is probing gas that is a projected
distance of $2 \mpc$ from the center of a system of galaxies with a density of
\nsdssg (\nsdss $\cmc$).  We have not accounted for thermal pressure in the
jets, as well as our inability to resolve their true width.  Additional
thermal pressure or smaller jets will necessitate a larger IGM
density to produce the observed jet geometry.  If we have overestimated the
radio galaxy speed or projection effects have led us to a smaller radius of
curvature then the necessary density will decrease.

The systematic errors reflect the range of IGM densities possible given the
range of jet speeds discussed in Section 3.  These speeds are measured by
assuming that flux asymmetries between the jet and counter-jet are the result
of relativistic beaming effects.  This beaming depends on the angle of the jets
from our line of sight as well as the jet speed.  By looking at large samples
of objects and assuming a uniform distribution of angles, a range of jet speeds
can be determined.  We lack the inclination information necessary to determine
jet speeds for individual sources.  \citet{2004MNRAS.351..727A} point out that
their distribution of jet speeds is asymmetric with the peak toward lower
speeds and that selection effects may be pushing the mean toward higher speeds.
Slower jet speeds would cause our IGM densities to tip towards the lower end of
the range in values.

The inclination of the source also affects the measured radius of curvature.
We use a value which is uncorrected for inclination to calculate the IGM
densities.  S1 appears relatively unprojected whereas S2 is less easily
characterized.  If we assume that a Hubble flow distance is accurate for S2,
and that it is traveling along the vector connecting its current position with
the cross indicating the center of the nearby system of galaxies, then its
inclination from the plane of the sky is $\sim 75^{o}$.  Deprojecting according
to this angle leads to a radius of curvature four times smaller and a
corresponding IGM density four times larger.  

We can compare the measured IGM densities to those seen in clusters and X-ray
bright groups of galaxies where the radial density profile of the hot gas in
the IGM is traced using X-ray data.  Gas densities at the center of clusters
are on the order of $10^{-25} \ \gcmc$ ($0.1 \cmc$) to $10^{-28} \gcmc$
($10^{-4} \cmc$) at a radius of $1 \mpc$ \citep{2006ApJ...640..691V}.  The IGM
density we measure near S1 is comparable to the density seen in X-ray
observations at a similar radius in the NGC 5044, NGC 533, and ESO 5520200
groups from \citet{2007ApJ...669..158G}.  Those groups have higher velocity
dispersions and may be dynamically older than the group containing S1.

Numerical simulations of the evolution of the IGM predict WHIM gas in the
local universe with a density distribution that is broadly peaked at
overdensities of $10-20$ and densities in filaments of $10-100$ relative to
the critical density ($1.0\times10^{-29} \gcmc$ for $H_0=73 \kmsmpc$)
\citep{2006ApJ...650..560C, 2006MNRAS.370..656D}.  The gas that S2 is tracing
has an overdensity of $\sim 90$.  Considering that this source is $2 \mpc$
from the center of a large group of galaxies it could be tracing gas in a
filament or gas that is only tenuously associated with the nearby group. 

Radio galaxies with bent jets in groups of galaxies provide 
evidence that intragroup gas exists in these systems even when the IGM is not
X-ray bright.  The densities derived in this paper indicate that groups of
galaxies and their surroundings are likely to contain significant reservoirs
of baryons.  Further work is underway with a larger number of sources and will
attempt to quantify the contribution of the intragroup medium to the baryon
density in the local universe. 

\acknowledgements EF would like to acknowledge the
support of a Wisconsin Space Grant Consortium Graduate Fellowship.  EW was
supported by NSF grant AST-0506628.  This research has made use of the
NASA/IPAC extragalactic database (NED) which is operated by the Jet Propulsion
Laboratory, Caltech, under contract with the National Aeronautics and Space
Administration.   We thank the staff of the GMRT who have made these
observations possible. GMRT is run by the National Centre for Radio
Astrophysics of the Tata Institute of Fundamental Research.

\clearpage

\begin{deluxetable*}{lccccccccc}
\tablecaption{Source Information\label{tab:sour}}
\tabletypesize{\scriptsize}
\tablehead{
\colhead{Source} &\colhead{$\alpha$} & \colhead{$\delta$} &\colhead{z} & \colhead{$L_{1440}$} & \colhead{$h$} & \colhead{$R$} & \colhead{$v_{gal}$} & \colhead{$P_{min,jet}$} & \colhead{$n_{\mbox{\tiny IGM}}$} \\
          &(J2000) & (J2000) & & $\mathrm{W\ Hz}^{-1}$ & (arcsec) & (arcsec) & ($\kms$) &  ($\mathrm{dynes~cm}^{-2}$) & ($\cmc$) }  
\startdata
S1   & 12 49 42.2  &  30 38 38 & 0.194 &$1.61\times10^{25}$   & 4.1 & 13$\pm$1  &$250_{-100}^{+20}$ & $2\times10^{-11}$ & \nfirst \\  
S2 & 08 53 31.9  &  23 24 00 & 0.306   &$1.88\times10^{25}$   & 5.8 & 23$\pm$2  &$570\pm60$         & $7\times10^{-12}$ & \nsdss   
\enddata
\tablecomments{$L_{1440}$ is from \citet{2001AJ....121.2915B}.}
\end{deluxetable*}

\clearpage

\begin{thebibliography}{38}
\expandafter\ifx\csname natexlab\endcsname\relax\def\natexlab#1{#1}\fi

\bibitem[{{Adelman-McCarthy et al.}(2008)}]{2008ApJS..175..297A}
{Adelman-McCarthy et al.} 2008, \apjs, 175, 297

\bibitem[{{Arnaud}(2004)}]{2004HEAD....8.1629A}
{Arnaud}, K. 2004, in Bulletin of the American Astronomical Society, Vol.~36,
  Bulletin of the American Astronomical Society, 934--+

\bibitem[{{Arshakian} \& {Longair}(2004)}]{2004MNRAS.351..727A}
{Arshakian}, T.~G. \& {Longair}, M.~S. 2004, \mnras, 351, 727

\bibitem[{{Becker} {et~al.}(1995){Becker}, {White}, \&
  {Helfand}}]{1995ApJ...450..559B}
{Becker}, R.~H., {White}, R.~L., \& {Helfand}, D.~J. 1995, \apj, 450, 559

\bibitem[{{Beers} {et~al.}(1990){Beers}, {Flynn}, \&
  {Gebhardt}}]{1990AJ....100...32B}
{Beers}, T.~C., {Flynn}, K., \& {Gebhardt}, K. 1990, \aj, 100, 32

\bibitem[{{Begelman} {et~al.}(1979){Begelman}, {Rees}, \&
  {Blandford}}]{1979Natur.279..770B}
{Begelman}, M.~C., {Rees}, M.~J., \& {Blandford}, R.~D. 1979, \nat, 279, 770

\bibitem[{{Blanton} {et~al.}(2001){Blanton}, {Gregg}, {Helfand}, {Becker}, \&
  {Leighly}}]{2001AJ....121.2915B}
{Blanton}, E.~L., {Gregg}, M.~D., {Helfand}, D.~J., {Becker}, R.~H., \&
  {Leighly}, K.~M. 2001, \aj, 121, 2915

\bibitem[{{Bridle} \& {Perley}(1984)}]{1984ARA&A..22..319B}
{Bridle}, A.~H. \& {Perley}, R.~A. 1984, \araa, 22, 319

\bibitem[{{Burns} {et~al.}(1987){Burns}, {Hanisch}, {White}, {Nelson},
  {Morrisette}, \& {Moody}}]{1987AJ.....94..587B}
{Burns}, J.~O., {Hanisch}, R.~J., {White}, R.~A., {Nelson}, E.~R.,
  {Morrisette}, K.~A., \& {Moody}, J.~W. 1987, \aj, 94, 587

\bibitem[{{Burns} \& {Owen}(1980)}]{1980AJ.....85..204B}
{Burns}, J.~O. \& {Owen}, F.~N. 1980, \aj, 85, 204

\bibitem[{{Cen} \& {Ostriker}(2006)}]{2006ApJ...650..560C}
{Cen}, R. \& {Ostriker}, J.~P. 2006, \apj, 650, 560

\bibitem[{{Dav{\'e}} {et~al.}(2001){Dav{\'e}}, {Cen}, {Ostriker}, {Bryan},
  {Hernquist}, {Katz}, {Weinberg}, {Norman}, \& {O'Shea}}]{2001ApJ...552..473D}
{Dav{\'e} et al.} 2001, \apj,
  552, 473

\bibitem[{{Dickey} \& {Lockman}(1990)}]{1990ARA&A..28..215D}
{Dickey}, J.~M. \& {Lockman}, F.~J. 1990, \araa, 28, 215

\bibitem[{{Doe} {et~al.}(1995){Doe}, {Ledlow}, {Burns}, \&
  {White}}]{1995AJ....110...46D}
{Doe}, S.~M., {Ledlow}, M.~J., {Burns}, J.~O., \& {White}, R.~A. 1995, \aj,
  110, 46

\bibitem[{{Dolag} {et~al.}(2006){Dolag}, {Meneghetti}, {Moscardini}, {Rasia},
  \& {Bonaldi}}]{2006MNRAS.370..656D}
{Dolag}, K., {Meneghetti}, M., {Moscardini}, L., {Rasia}, E., \& {Bonaldi}, A.
  2006, \mnras, 370, 656

\bibitem[{{Ekers} {et~al.}(1978){Ekers}, {Fanti}, {Lari}, \&
  {Ulrich}}]{1978A&A....69..253E}
{Ekers}, R.~D., {Fanti}, R., {Lari}, C., \& {Ulrich}, M.-H. 1978, \aap, 69, 253

\bibitem[{{Fukugita} {et~al.}(1998){Fukugita}, {Hogan}, \&
  {Peebles}}]{1998ApJ...503..518F}
{Fukugita}, M., {Hogan}, C.~J., \& {Peebles}, P.~J.~E. 1998, \apj, 503, 518

\bibitem[{{Fukugita} \& {Peebles}(2004)}]{2004ApJ...616..643F}
{Fukugita}, M. \& {Peebles}, P.~J.~E. 2004, \apj, 616, 643

\bibitem[{{Gastaldello} {et~al.}(2007){Gastaldello}, {Buote}, {Humphrey},
  {Zappacosta}, {Bullock}, {Brighenti}, \& {Mathews}}]{2007ApJ...669..158G}
{Gastaldello}, F., {Buote}, D.~A., {Humphrey}, P.~J., {Zappacosta}, L.,
  {Bullock}, J.~S., {Brighenti}, F., \& {Mathews}, W.~G. 2007, \apj, 669, 158

\bibitem[{{G{\'o}mez} {et~al.}(2003){G{\'o}mez}, {Nichol}, {Miller}, {Balogh},
  {Goto}, {Zabludoff}, {Romer}, {Bernardi}, {Sheth}, {Hopkins}, {Castander},
  {Connolly}, {Schneider}, {Brinkmann}, {Lamb}, {SubbaRao}, \&
  {York}}]{2003ApJ...584..210G}
{G{\'o}mez et al.} 2003, \apj,
  584, 210

\bibitem[{{Goto} {et~al.}(2003){Goto}, {Yamauchi}, {Fujita}, {Okamura},
  {Sekiguchi}, {Smail}, {Bernardi}, \& {Gomez}}]{2003MNRAS.346..601G}
{Goto}, T., {Yamauchi}, C., {Fujita}, Y., {Okamura}, S., {Sekiguchi}, M.,
  {Smail}, I., {Bernardi}, M., \& {Gomez}, P.~L. 2003, \mnras, 346, 601

\bibitem[{{Jetha} {et~al.}(2006){Jetha}, {Hardcastle}, \&
  {Sakelliou}}]{2006MNRAS.368..609J}
{Jetha}, N.~N., {Hardcastle}, M.~J., \& {Sakelliou}, I. 2006, \mnras, 368, 609

\bibitem[{{Jones} \& {Owen}(1979)}]{1979ApJ...234..818J}
{Jones}, T.~W. \& {Owen}, F.~N. 1979, \apj, 234, 818

\bibitem[{{Kantharia} {et~al.}(2005){Kantharia}, {Ananthakrishnan},
  {Nityananda}, \& {Hota}}]{2005A&A...435..483K}
{Kantharia}, N.~G., {Ananthakrishnan}, S., {Nityananda}, R., \& {Hota}, A.
  2005, \aap, 435, 483

\bibitem[{{Kurtz} \& {Mink}(1998)}]{1998PASP..110..934K}
{Kurtz}, M.~J. \& {Mink}, D.~J. 1998, \pasp, 110, 934

\bibitem[{{McConnachie} {et~al.}(2007){McConnachie}, {Venn}, {Irwin}, {Young},
  \& {Geehan}}]{2007ApJ...671L..33M}
{McConnachie}, A.~W., {Venn}, K.~A., {Irwin}, M.~J., {Young}, L.~M., \&
  {Geehan}, J.~J. 2007, \apjl, 671, L33

\bibitem[{{Mulchaey}(2000)}]{2000ARA&A..38..289M}
{Mulchaey}, J.~S. 2000, \araa, 38, 289

\bibitem[{{O'Dea} \& {Owen}(1987)}]{1987ApJ...316...95O}
{O'Dea}, C.~P. \& {Owen}, F.~N. 1987, \apj, 316, 95

\bibitem[{{Pacholczyk}(1970)}]{1970ranp.book.....P}
{Pacholczyk}, A.~G. 1970, {Radio astrophysics. Nonthermal processes in galactic
  and extragalactic sources} (Series of Books in Astronomy and Astrophysics,
  San Francisco: Freeman, 1970)

\bibitem[{{Pisano} {et~al.}(2004){Pisano}, {Wakker}, {Wilcots}, \&
  {Fabian}}]{2004AJ....127..199P}
{Pisano}, D.~J., {Wakker}, B.~P., {Wilcots}, E.~M., \& {Fabian}, D. 2004, \aj,
  127, 199

\bibitem[{{Rasmussen} {et~al.}(2006){Rasmussen}, {Ponman}, \&
  {Mulchaey}}]{2006MNRAS.370..453R}
{Rasmussen}, J., {Ponman}, T.~J., \& {Mulchaey}, J.~S. 2006, \mnras, 370, 453

\bibitem[{{Savage} {et~al.}(2003){Savage}, {Sembach}, {Wakker}, {Richter},
  {Meade}, {Jenkins}, {Shull}, {Moos}, \& {Sonneborn}}]{2003ApJS..146..125S}
{Savage}, B.~D., {Sembach}, K.~R., {Wakker}, B.~P., {Richter}, P., {Meade}, M.,
  {Jenkins}, E.~B., {Shull}, J.~M., {Moos}, H.~W., \& {Sonneborn}, G. 2003,
  \apjs, 146, 125

\bibitem[{{Sivakoff} {et~al.}(2004){Sivakoff}, {Sarazin}, \&
  {Carlin}}]{2004ApJ...617..262S}
{Sivakoff}, G.~R., {Sarazin}, C.~L., \& {Carlin}, J.~L. 2004, \apj, 617, 262

\bibitem[{{Tripp} {et~al.}(1998){Tripp}, {Lu}, \&
  {Savage}}]{1998ASPC..143..261T}
{Tripp}, T.~M., {Lu}, L., \& {Savage}, B.~D. 1998, in ASP Conf. Ser. 143: The
  Scientific Impact of the Goddard High Resolution Spectrograph, ed. J.~C.
  {Brandt}, T.~B. {Ake}, \& C.~C. {Petersen}, 261--+

\bibitem[{{Tripp} {et~al.}(2000){Tripp}, {Savage}, \&
  {Jenkins}}]{2000ApJ...534L...1T}
{Tripp}, T.~M., {Savage}, B.~D., \& {Jenkins}, E.~B. 2000, \apjl, 534, L1

\bibitem[{{Venkatesan} {et~al.}(1994){Venkatesan}, {Batuski}, {Hanisch}, \&
  {Burns}}]{1994ApJ...436...67V}
{Venkatesan}, T.~C.~A., {Batuski}, D.~J., {Hanisch}, R.~J., \& {Burns}, J.~O.
  1994, \apj, 436, 67

\bibitem[{{Vikhlinin} {et~al.}(2006){Vikhlinin}, {Kravtsov}, {Forman}, {Jones},
  {Markevitch}, {Murray}, \& {Van Speybroeck}}]{2006ApJ...640..691V}
{Vikhlinin}, A., {Kravtsov}, A., {Forman}, W., {Jones}, C., {Markevitch}, M.,
  {Murray}, S.~S., \& {Van Speybroeck}, L. 2006, \apj, 640, 691

\bibitem[{{Worrall} \& {Birkinshaw}(2000)}]{2000ApJ...530..719W}
{Worrall}, D.~M. \& {Birkinshaw}, M. 2000, \apj, 530, 719

\end{thebibliography}
\end{document}